\documentclass{emulateapj}
\newcommand{\of}{0509$-$67.5}
\newcommand{\ch}{\textit{Chandra}}
\newcommand{\xmm}{\textit{XMM-Newton}}

\slugcomment{Draft Version \today}

\shorttitle{SNR \of\ AND ITS PARENT SUPERNOVA}
\shortauthors{Badenes et al.}

\begin{document}

\title{THE PERSISTENCE OF MEMORY, OR HOW THE X-RAY SPECTRUM OF SNR 0509$-$67.5 REVEALS THE BRIGHTNESS OF ITS PARENT TYPE Ia SUPERNOVA}

\author{Carles Badenes\altaffilmark{1,2,3}, John P. Hughes\altaffilmark{1,2}, Gamil Cassam-Chena\"{i}\altaffilmark{1}, and Eduardo Bravo\altaffilmark{4}}

\altaffiltext{1}{Department of Astrophysical Sciences, Princeton University, Peyton Hall, Ivy lane, Princeton, NJ 08544-1001; badenes@astro.princeton.edu}

\altaffiltext{2}{Department of Physics and Astronomy, Rutgers University, 136 Frelinghuysen Rd., Piscataway NJ
  08854-8019; jph@physics.rutgers.edu, chenai@physics.rutgers.edu}

\altaffiltext{3}{\textit{Chandra} Fellow}

\altaffiltext{4}{Departament de F\'{i}sica i Enginyeria Nuclear, Universitat Polit\`{e}cnica de Catalunya, Diagonal 647,
  Barcelona 08028, Spain; and Institut d'Estudis Espacials de Catalunya, Campus UAB, Facultat de Ci\`{e}ncies. Torre
  C5. Bellaterra, Barcelona 08193, Spain; eduardo.bravo@upc.es}

\begin{abstract}
  We examine the dynamics and X-ray spectrum of the young Type Ia supernova remnant \of\ in the context of the recent
  results obtained from the optical spectroscopy of its light echo. Our goal is to estimate the kinetic energy of the
  supernova explosion using \ch\ and \xmm\ observations of the supernova remnant, thus placing the birth event of \of\
  in the sequence of dim to bright Type Ia supernovae. We base our analysis on a standard grid of one-dimensional
  delayed detonation explosion models, together with hydrodynamic and X-ray spectral calculations of the supernova
  remnant evolution. From the remnant dynamics and the properties of the O, Si, S, and Fe emission in its X-ray spectrum
  we conclude that \of\ was originated $\sim 400$ years ago by a bright, highly energetic Type Ia explosion similar to
  SN 1991T. Our best model has a kinetic energy of $1.4 \times 10^{51}$ erg and synthesizes $0.97 \, \mathrm{M_{\odot}}$
  of $^{56}$Ni. These results are in excellent agreement with the age estimate and spectroscopy from the light echo. We
  have thus established the first connection between a Type Ia supernova and its supernova remnant based on a detailed
  quantitative analysis of both objects.
\end{abstract}


\section{INTRODUCTION}
\label{sec:Intro}

Astronomical observations can probe the material ejected by supernova (SN) explosions during two transient phases with
very different time scales. The initial optical transient (the SN itself) lasts for several months, and the ejecta
structure can be studied through the emission and absorption lines produced as the photosphere recedes into the exploded
star. After the SN fades away, the ejected material starts to interact with its surroundings, and the supernova remnant
(SNR) phase begins. In this phase, the ejecta structure is revealed by the X-ray emission of the material heated by the
reverse shock on timescales of hundreds or thousands of years. Both approaches are valid in principle, but up to now the
disparity of the timescales involved has made it impossible to verify their mutual consistency by applying them to the
same object.

The serendipitous discovery of light echoes associated with the young SNRs \of, 0519$-$69.0, and N103B in the Large
Magellanic Cloud (LMC) by \citet{rest05:LMC_light_echoes} (henceforth R05) has opened new possibilities for establishing
connections between SNRs and their parent SNe. In particular, spectroscopy of the light echoes has the potential to
confirm the type of the SN explosion in a straightforward way, avoiding the difficulties inherent to typing SNRs from
their X-ray spectra \citep[for an in-depth discussion of these difficulties, see][]{rakowski05:G337}. In a recent paper,
\citet{rest07:0509_light_echo} (henceforth R07) examined the light echo associated with SNR \of, and established that
this object originated in a Type Ia explosion, in agreement with the longstanding claims based on X-ray observations
\citep[][henceforth WH04]{hughes95:typing_SN_from_SNR,warren03:0509-67.5}. Furthermore, the quality of the light echo
was such that it allowed for a meaningful comparison with the spectra of a variety of Type Ia SNe. Based on these
comparisons, R07 concluded that the parent event of SNR \of\ belonged to the group of overluminous, highly energetic
Type Ia SNe whose prototype is SN 1991T \citep{benetti05:Ia_diversity}.

These results offer an excellent opportunity to revisit the X-ray spectrum of SNR \of. Recent developments in modeling
the thermal emission in young Type Ia SNRs using hydrodynamic calculations and nonequilibrium ionization simulations
\citep[HD+NEI models,][]{badenes03:xray,badenes05:xray} have led to a better understanding of the relationship between
the structure of SN ejecta and the X-ray spectra of SNRs. By applying HD+NEI models to the Tycho SNR,
\citet{badenes06:tycho} were able to estimate the kinetic energy and nucleosynthetic yields of the explosion. We can now
do the same for \of, comparing the results of the HD+NEI models with the light echo spectroscopy. In particular, we want
to examine to what extent is it possible to determine the brightness of a Type Ia SN by studying the X-ray emission from
its SNR hundreds of years after the explosion.

This paper is organized as follows. In \S~\ref{sec:DDT}, we begin with a brief overview of delayed detonation models. In
\S~\ref{sec:Dynamics}, we use these explosion models, together with the forward shock radius and velocity to constrain
the dynamics of SNR \of. In \S~\ref{sec:Observations}, we review the \ch\ and \xmm\ observations of \of, and we
calculate the values for the most relevant line centroids and line flux ratios. In \S~\ref{sec:Modeling} we describe the
HD+NEI models, and we evaluate them against both the dynamic and spectral constraints posed by the observations of
\of. Finally, we discuss our results and present our conclusions in \S~\ref{sec:Disc-Concl}.

\section{OVERVIEW OF DELAYED DETONATION MODELS}
\label{sec:DDT}

Since they were first introduced by \citet{khokhlov91:ddt}, delayed detonations (DDTs) have become the most successful
models for Type Ia SNe. In this kind of explosion, the burning front starts propagating as a subsonic flame
(deflagration) in the central regions of a C+O white dwarf (WD) close to the Chandrasekhar mass, and then a transition
to a supersonic regime (detonation) is artificially induced, usually at a prescribed density $\rho_{tr}$. The resulting
ejecta structure is characterized by an approximately exponential density profile, with a composition dominated by
Fe-peak nuclei in the center, surrounded by a shell rich in intermediate mass elements (IMEs: Si, S, Ar, Ca, etc.), and
a thinner outer region dominated by O. Minor traces of C might remain at large radii, but most of the WD is burnt in the
explosion. Spherically symmetric DDT models are able to explain the fundamental properties of Type Ia SNe, including
light curve shapes \citep{hoeflich96:nosubCh}, optical and IR spectral evolution
\citep{wheeler98:TypeIa-diagnostics-IR,baron06:SNIa_Maximum_Light,marion06:lowCinSNIa,gerardy07:SNIa_midIR}, and
spectropolarimetric observations \citep{wang07:SNIa_spectropolarimetry}.

\begin{figure}

  \centering
 
  \includegraphics[angle=90,scale=0.9]{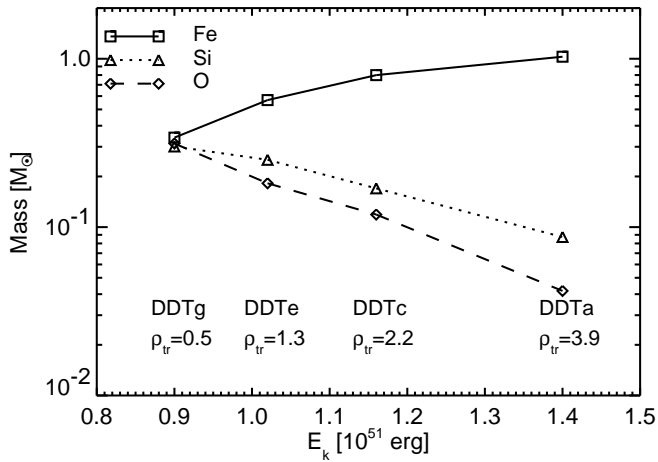}

  \caption{Nucleosynthetic yields of Fe (after radioactive decays, solid line, squares), Si (dotted line, triangles),
    and O (dashed line, diamonds) in the DDT models as a function of $E_{k}$. Each model is labeled in the plot,
    indicating the corresponding value of $\rho_{tr}$ in units of $10^{7}\,\mathrm{g\,cm^{-3}}$. \label{fig-1}}

\end{figure}

The deflagration-to-detonation transition density $\rho_{tr}$ is the most important parameter in DDT explosions. Models
with higher values of $\rho_{tr}$ are more energetic, and synthesize more Fe-peak elements, less IMEs, and less O than
models with low values of $\rho_{tr}$. Since Type Ia light curves are powered by the radioactive decay of $^{56}$Ni,
more energetic models lead to more luminous SNe \citep{stritzinger06:consistent_estimates_56Ni_Ias}. The luminosity of
Type Ia SNe is also tightly correlated to the light curve width \citep{phillips93:Ia-LLCcorrelation}, although the
physical processes connecting this parameter and the mass of $^{56}$Ni synthesized in the explosion are not as simple as
was initially thought \citep[see][]{kasen07:wlr,woosley07:light_curves}. In any case, the relationship between Type Ia
SN luminosity (or light curve width) and the chemical structure of SN ejecta has now been firmly established by
\citet{mazzali07:zorro}, and is consistent with the predictions of one-dimensional DDT models. 

The fundamental properties of the grid of DDT models that we will use in the present work are plotted in Figure
\ref{fig-1}, which can be compared to the `Zorro' diagram in \citet{mazzali07:zorro}. Models DDTa, DDTc, and DDTe are
taken from \citet{badenes03:xray,badenes05:xray}; model DDTg has been calculated with the same code
\citep[see][]{bravo96:code,badenes03:xray}, and extends the grid to lower values of $E_{k}$. The total masses of
$^{56}$Ni synthesized in the models are $0.97\,\mathrm{M_{\odot}}$ (DDTa), $0.74\,\mathrm{M_{\odot}}$ (DDTc),
$0.51\,\mathrm{M_{\odot}}$ (DDTe), $0.29\,\mathrm{M_{\odot}}$ (DDTg). To put these values in context, the estimated
$^{56}$Ni masses in the 20 objects considered `normal' by \citet{mazzali07:zorro} range between
$0.94\pm0.05\,\mathrm{M_{\odot}}$ (for SN 1991T) and $0.24\pm0.05\,\mathrm{M_{\odot}}$ (for SN 1991M). A theoretical
upper limit to the mass of radioactive Ni that can be obtained from the thermonuclear burning of a Chandrasekhar C+O WD
is set by the prompt detonation model DET in \citet{badenes03:xray}, which yields $1.16\,\mathrm{M_{\odot}}$ of
$^{56}$Ni with $E_{k}= 1.6 \times 10^{51}$ erg.

\section{SNR DYNAMICS}
\label{sec:Dynamics}

The dynamics of \of\ are constrained by two pieces of observational evidence. The first is the angular radius measured
by \ch\ \citep[$15.1"$, see Table 3 in ][]{badenes07:outflows}, which translates into a linear radius of
$1.1\times10^{19}$ cm ($3.7$ pc) at the known distance to the LMC \citep[50 kpc,][]{alves04:LMC_Distance}. This
measurement can be considered very accurate, because the errors on both angular radius and distance are extremely small
(at the few percent level). The second observable is the forward shock (FS) velocity inferred by modeling the broad
component of the Ly$\beta$ emission line, $3600-7100\,\mathrm{km \, s^{-1}}$ \citep{ghavamian07:FUSE_LMC_SNRs}.  Being a
less direct measurement, the FS velocity is more subject to uncertainty than the FS radius, so this value should be
considered with some measure of caution \citep[see discussion in \S~5.1 of][]{badenes07:outflows}. To reproduce the
dynamics of \of, we have used a one-dimensional hydrodynamic code to simulate the interaction between the ejecta density
profiles from the DDT explosion models in our grid and a uniform ambient medium \citep[AM; for more details on the
models and a justification of the uniform AM hypothesis, see][]{badenes07:outflows}. A successful SNR model must be able
to match both FS radius and velocity for a reasonable combination of age $t$ and AM density $\rho_{AM}$.

\begin{figure}

  \centering
 
  \includegraphics[angle=90,scale=0.9]{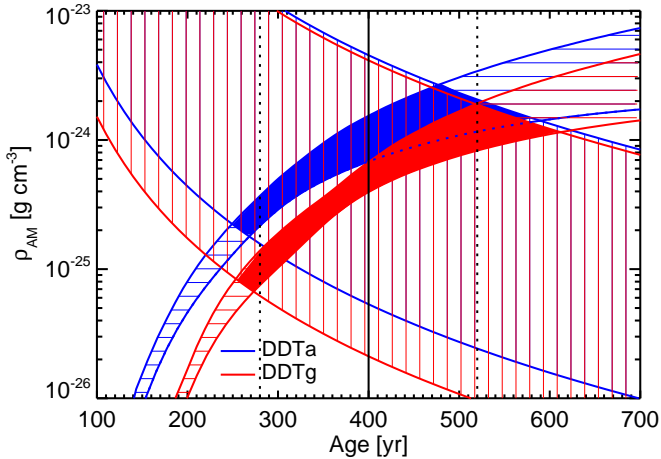}

  \caption{Constraints on the dynamics of \of\ from the observed FS radius and velocity. The horizontally striped areas
    indicate the regions of the parameter space where the FS radius lies between $R_{FS}$ and $1.1 \times R_{CD}$ in
    models DDTa (blue) and DDTg (red). The vertically striped areas indicate the regions of the parameter space where
    the FS velocity lies between $u_{FS}$ and $u_{CD}$ in each model. The areas where both constraints overlap are
    highlighted with solid colors. The age estimate from the light echoes is represented with the vertical solid and
    dotted lines. \label{fig-2}}

\end{figure}

The comparison between hydrodynamic models and observations, however, is not completely straightforward, because our
simulations do not include the effect of cosmic ray (CR) acceleration. This physical process can strongly modify the
dynamics of the SNR, slowing down the FS and reducing the gap between FS and contact discontinuity (CD) by an amount
that depends on the acceleration efficiency \citep{ellison04:hd+cr}. Indeed, WH04 found evidence for a significant
nonthermal component in the X-ray spectrum of \of\ and argued that CR acceleration was taking place at the FS. Although
the presence of nonthermal emission does not necessarily imply that the dynamics are CR-modified, we consider this to be
a likely possibility. It is important to note that CR acceleration at the FS does not have an impact on the dynamics of
the shocked ejecta, and in particular, it does not affect the position of the CD \citep[at least to first order, see
Fig. 2 in][]{ellison07:CR_Acceleration_Thermal_Nonthermal}. This means that the CR-modified SNR radius must lie between
the FS radius $R_{FS}$ and the CD radius $R_{CD}$ in hydrodynamic models without CR acceleration. For one-dimensional
models, it is necessary to allow for the fact that Rayleigh-Taylor instabilities at the CD increase the projected radius
by $\sim 10 \%$ \citep{wang01:Ia-inst-clump}. Likewise, the CR-modified FS velocity must lie between the FS velocity
$u_{FS}$ and the CD velocity $u_{CD}$. We have summarized these constraints in Figure \ref{fig-2} for the most and least
energetic models in our DDT grid, DDTa and DDTg. The DDT models require that the SNR age be between 250 and 610 yr to
reproduce the FS dynamics, in very good agreement with the completely independent $400 \pm 120$ yr estimate from the
light echoes (R05). The allowed values of $\rho_{AM}$ lie between $5\times10^{-26}$ and $2\times10^{-24}\,\mathrm{g \,
  cm^{-3}}$, with more energetic models demanding higher AM densities. The FS radius provides the strongest dynamical
constraints, imposing a tight correlation between $\rho_{AM}$ and age that we shall revisit in
\S~\ref{sub:DDT_Otherages}.

\section{X-RAY OBSERVATIONS}
\label{sec:Observations}

\subsection{\ch\  and \xmm\ Data Sets}
\label{sub:Data-Sets}

SNR \of\ has been observed by the CCD cameras on both \ch\ and \xmm. The most detailed study of its X-ray emission to
date was presented by WH04, who analyzed a \ch\ ACIS-S data set taken on 2000 May 12-13 (ObsID 776, PI J.Hughes). The
net exposure time was 47.9 ks, and the image was taken on the back-illuminated S3 chip. The authors found that the X-ray
spectrum is dominated by line emission from the SN ejecta, with a small continuum contribution, probably nonthermal
emission from the FS. The centroid of the Si K$\alpha$ line blend indicates an unusually low (below He-like) ionization
state for the plasma. The abundances obtained from plane-parallel shock fits favor a Type Ia origin, in agreement with
the previous qualitative analysis of \citet{hughes95:typing_SN_from_SNR}. These abundances were compared to the yields
of two type Ia SN explosion models from \citet{iwamoto99:SNIa-model-grid}, a `fast deflagration' and a DDT, with the
fitted values showing some preference for the DDT model, albeit with gross discrepancies in the Fe abundance that were
noted by the authors. This might be due to the simplicity of the plane shock models used by WH04, but the poor
statistics of the Fe K$\alpha$ blend in the \ch\ ACIS-S data set also make it difficult to constrain the properties of
shocked Fe in \of.

For the present work, we have recalibrated the \ch\ ACIS-S data using the latest versions of CIAO (3.4) and CALDB
(3.4.0), which yielded 48.5 ks of useful exposure. To complement the \ch\ data set, we have also reduced and analyzed
the \xmm\ observation taken on 2000 July 4 (ObsID 0111130201, PI M.Watson). The instrument modes in this observation
were small window for EPIC-MOS1 (frame time: 300 ms), large window for EPIC-MOS2 (900 ms) and EPIC-pn (48 ms). The
medium filter was used. We reduced the data using the latest SAS version (7.1.0) and calibration files. We applied the
flare rejection procedure described in \citet{cassam-chenai04:RX_J1713.7-3946}, which left a total exposure of 32.6 ks
for all cameras. To create spectra, we selected single and double events (pattern $\leq 4$) for the EPIC-pn camera,
which greatly improved the statistics in the Fe K$\alpha$ line. The spectra were then rebinned to achieve a
signal-to-noise ratio $>3\sigma$.

\begin{figure}

  \centering
 
  \includegraphics[angle=90,scale=0.9]{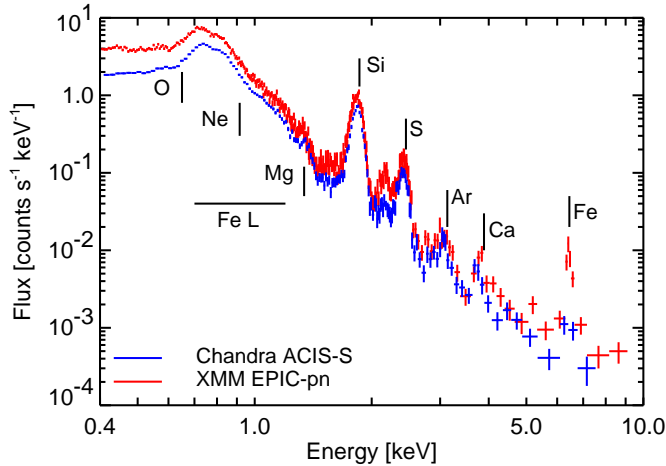}

  \caption{Spatially integrated spectrum from \ch\ ACIS-S (blue) and \xmm\ EPIC-pn (red). The expected centroids of the
    K$\alpha$ blend from the He-like ions of O, Ne, Mg, Si, S, Ar, and Ca are indicated for reference, as is the Fe
    K$\alpha$ blend and the region dominated by Fe L-shell lines. \label{fig-3}}

\end{figure}

A preliminary inspection of the calibrated and reduced spectra reveals that the centroids of the brightest lines in the
EPIC-MOS cameras appear shifted towards high energies compared to both EPIC-pn and \ch\ ACIS-S. The Si K$\alpha$
centroid values obtained with the different CCD instruments, with standard $90\%$ confidence intervals, are $1.850 \pm
0.002$ keV (EPIC-MOS1), $1.847^{+0.003}_{-0.002}$ keV (EPIC-MOS2), $1.834 \pm 0.002$ keV (EPIC-pn), and
$1.833^{+0.003}_{-0.002}$ keV (ACIS-S). The overlap between the ACIS-S and EPIC-pn values is important, because the
energy scale of the \ch\ ACIS-S data set was verified using on-board calibration sources (see \S~2 in
WH04). Furthermore, EPIC-pn has a substantially larger effective area at the Fe K$\alpha$ line than EPIC-MOS. In view
of this, we have chosen to maximize the consistency of our data sets by not including the EPIC-MOS spectra in our
analysis. The spatially integrated \ch\ ACIS-S and \xmm\ EPIC-pn spectra are plotted in Fig.~\ref{fig-3}.

\subsection{Spectral Fits to the Line Emission}
\label{sub:Spectral-Fits}

\begin{figure*}

  \centering
 
  \includegraphics[angle=90,scale=0.8]{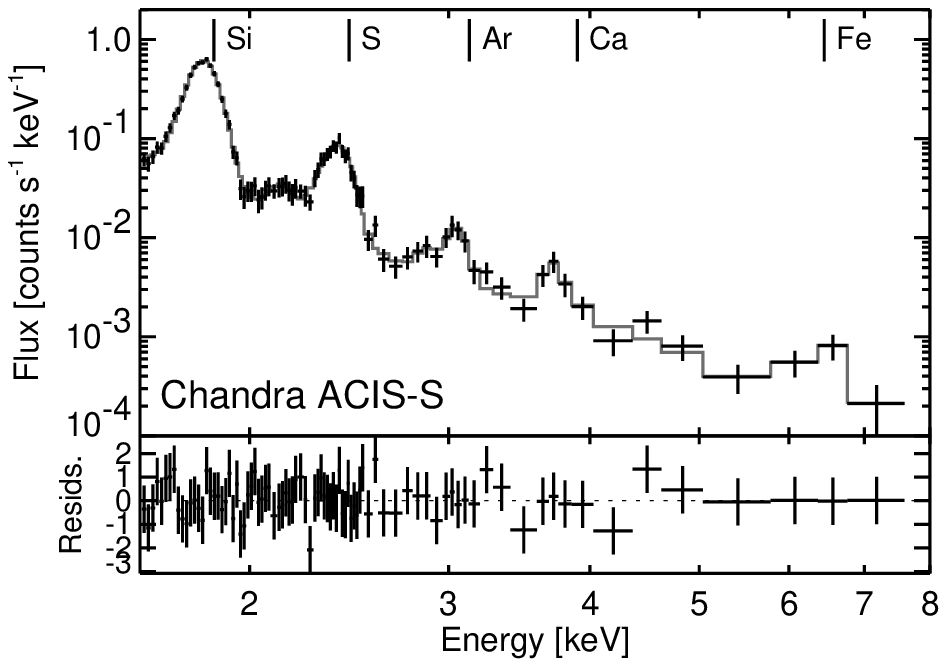}
  \includegraphics[angle=90,scale=0.8]{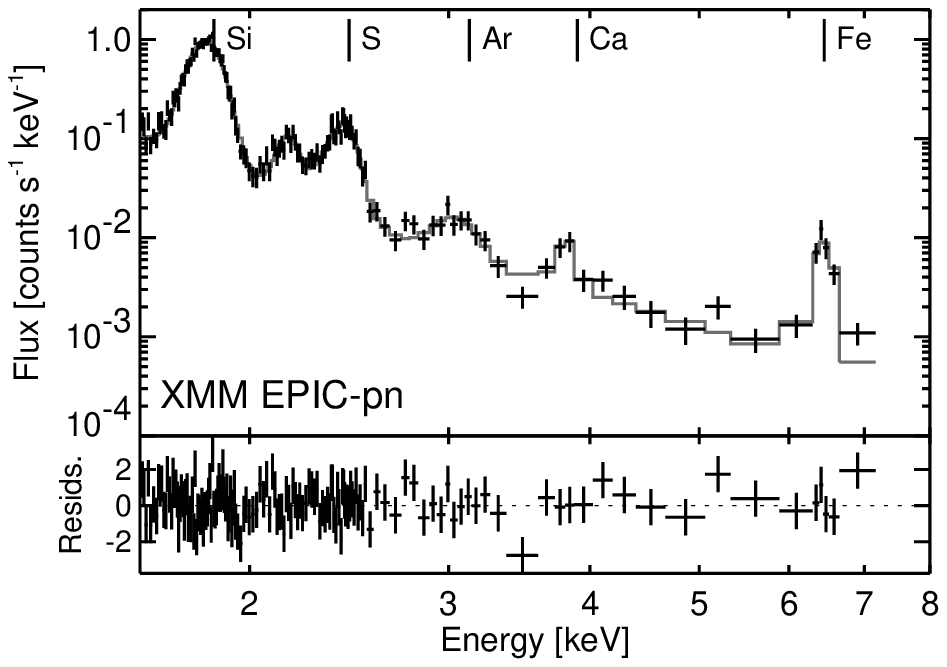}

  \caption{Spectral fits to the line emission in \of\ between 1.6 and 8.0 keV for \ch\ ACIS-S (left) and \xmm\ EPIC-pn
    (right). The expected centroids of the K$\alpha$ blends from the He-like ions of Si, S, Ar, and Ca, and the position
    of the Fe K$\alpha$ blend are indicated for reference. \label{fig-4}}

\end{figure*}

We have fitted the \ch\ ACIS-S and \xmm\ EPIC-pn data between 1.6 and 8.0 keV with a spectral model consisting of a
power law continuum and 16 Gaussian lines, including K-shell transitions from principal quantum level $n=2$ (K$\alpha$
blends) for Si, S, Ar, Ca, and Fe; K-shell transitions from $n=3,4,5$ (K$\beta$, K$\gamma$, and K$\delta$) for Si, S,
and Ar, and the Ly$\alpha$ lines for the H-like ions of Si and S. Only the centroids and fluxes of the K$\alpha$ blends
and the Si K$\beta$ line have been fitted. The centroids of the other K$\beta$ lines and all the K$\gamma$, K$\delta$,
and Ly$\alpha$ lines have been kept fixed at the nominal energies for He-like ions. The fluxes of the K$\gamma$ and
K$\delta$ lines have been tied to the respective K$\beta$ lines taking the ratios at $T_{e}=5 \times 10^{7}$ K from the
ATOMDB data base \citep{smith01:H_like_and_He_like_ions}. The resulting fits are shown on Figure \ref{fig-4} and the
fitted parameters are listed in Table \ref{tab-1}.

Outside the energy range of the fits, we have calculated a conservative upper limit to the O K$\alpha$ flux by assuming
that all the photons between 0.55 and 0.58 keV come from this blend, with interstellar absorption set at the maximum
value found by WH04, $N_{H}=0.076 \times 10^{22}\,\mathrm{cm^{2}}$. We note that the presence of O was necessary in the
plane shock fits performed by WH04, and that O lines are clearly seen in the grating observations (Hughes et al. in
preparation), but for this element it is especially hard to disentangle the ejecta contribution from the shocked AM
contribution (if any). Thus, we settle for the conservative upper limits listed in Table \ref{tab-1} to constrain our
ejecta models.  We have also calculated the flux in the broad band from 0.8 keV to 1.2 keV, which is dominated by
L-shell lines from Fe, with some contribution from K-shell Ne lines, assuming a nominal value of $N_{H}=0.07 \times
10^{22}\,\mathrm{cm^{2}}$. The principal line flux ratios normalized with respect to Si K$\alpha$ are listed in Table
\ref{tab-2}.

\begin{figure*}

  \centering
 
  \includegraphics[scale=0.6]{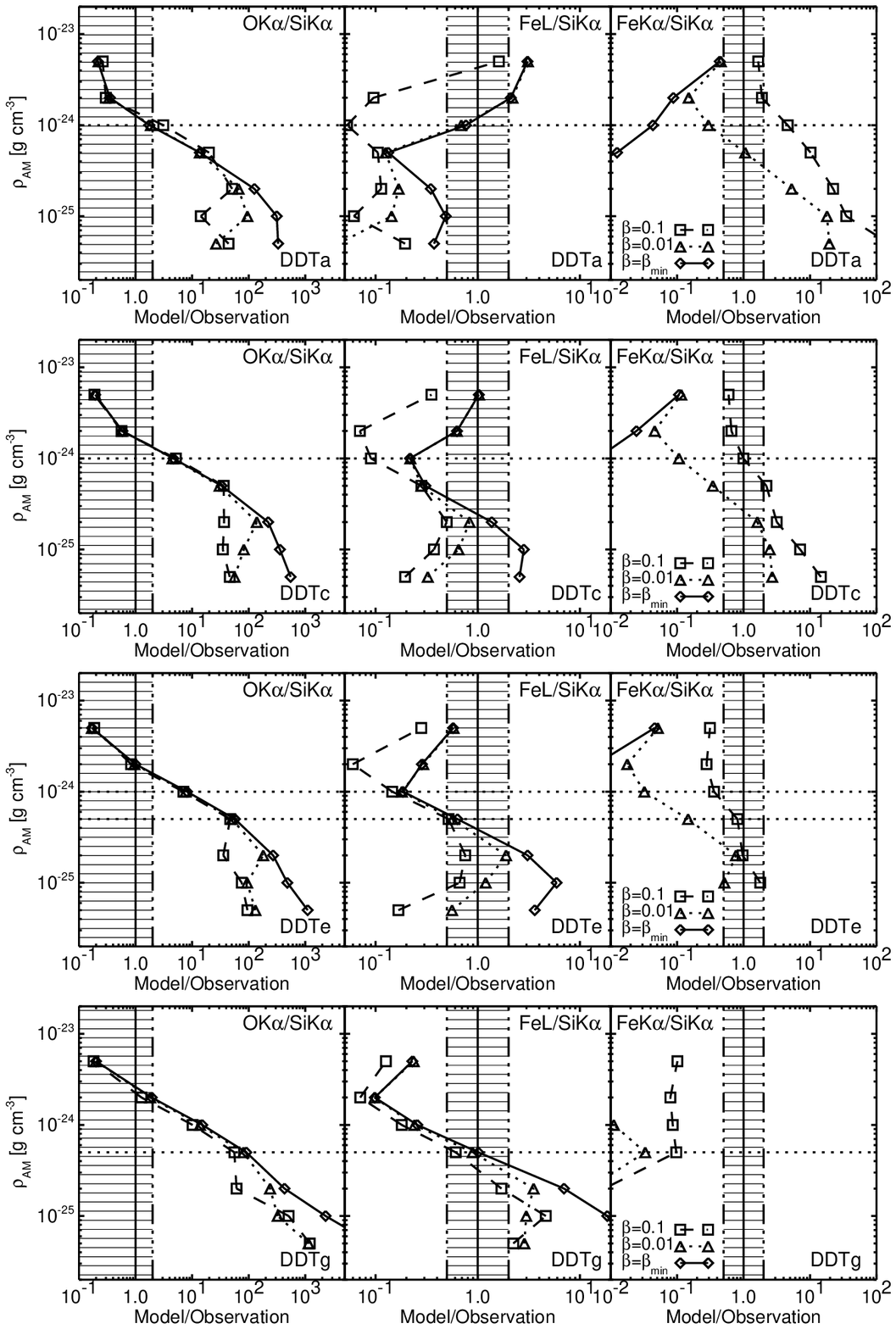}
  \includegraphics[scale=0.6]{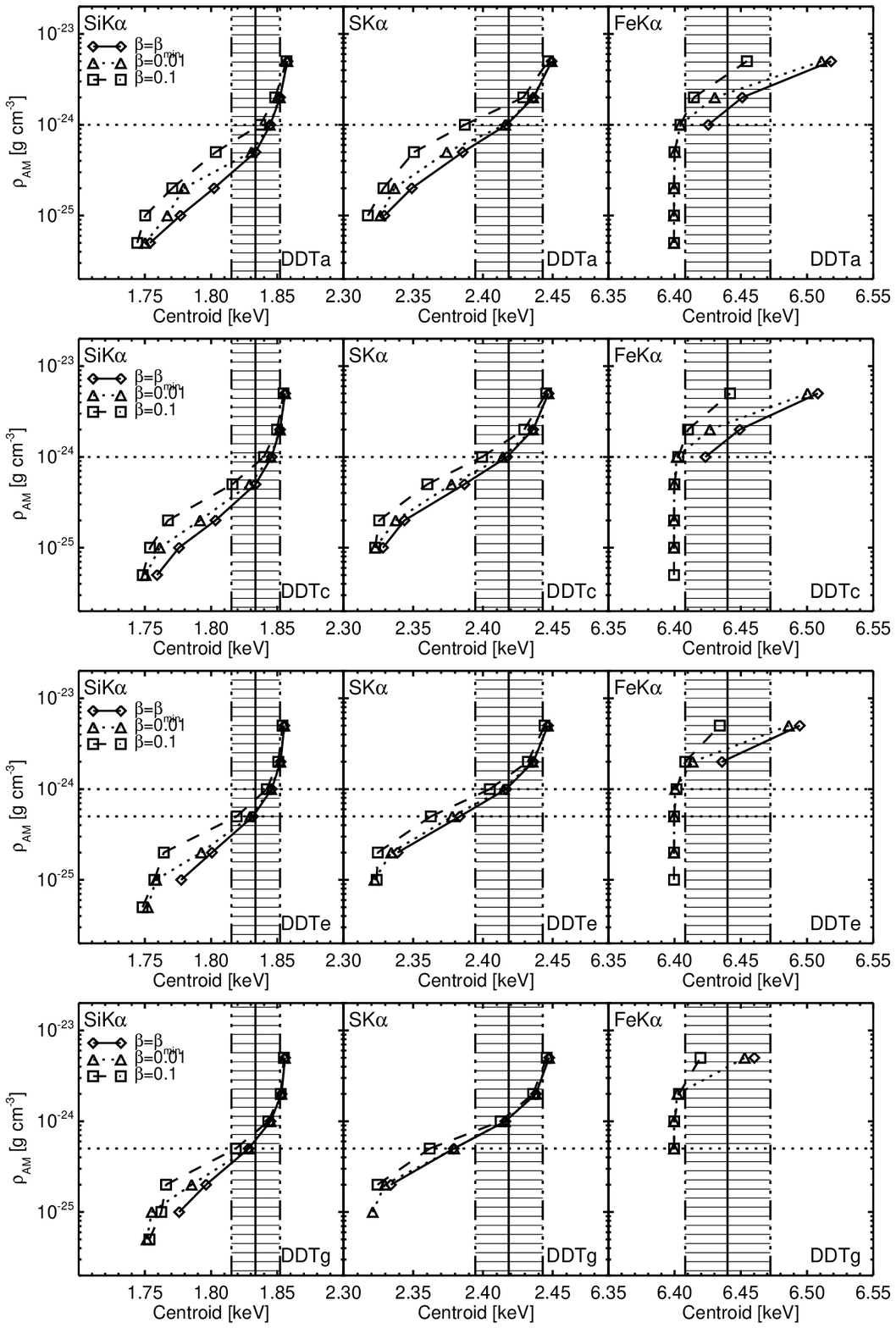}

  \caption{Fundamental line flux ratios (left panels) and line centroids (right panels) as a function of $\rho_{AM}$ in
    the DDT models at $t=400$ yr. For each model, the different values of $\beta$ are plotted with solid lines and
    diamonds ($\beta_{min}$), dotted lines and triangles ($\beta=0.01$), and dashed lines and squares ($\beta=0.1$). The
    line flux ratios are normalized to the observed values (from Table \ref{tab-2}), so that 1.0 is a perfect
    match. Tolerance ranges for line ratios and centroids are indicated by the striped regions around the observed
    values (vertical solid lines). The horizontal dotted lines mark the values of $\rho_{AM}$ that satisfy the dynamic
    constraints at t=400 yr, as discussed in \S~\ref{sec:Dynamics}. \label{fig-5}}

\end{figure*}

\begin{deluxetable}{lcc}
  \tablewidth{0pt}
  \tabletypesize{\scriptsize}
  \tablecaption{Spectral parameters in the SNR \of \label{tab-1}}
  \tablecolumns{3}
  \tablehead{
    \colhead{} &
    \colhead{\ch} &
    \colhead{\xmm}\\
    \colhead{Parameter} &
    \colhead{ACIS-S} &
    \colhead{EPIC-pn}
  }

  \startdata

  \cutinhead{Power Law Continuum}
  $\alpha$ &  $3.37^{+0.23}_{-0.21}$ &  $3.42^{+0.20}_{-0.19} $ \\
  Norm.  &  & \\
  $(10^{-6}\mathrm{photons\, cm^{-2}\, s^{-1}})$ & $416^{+89}_{-74}$ & $348^{+67}_{-57}$\\

  \cutinhead{Line Fluxes $(10^{-6}\mathrm{phot\, cm^{-2}\, s^{-1}})$}
  O K$\alpha$ \tablenotemark{a} & $<396$ & $<362$ \\
  Fe L \tablenotemark{b} & $1177$ & $958$ \\
  Si K$\alpha$ & $123 \pm 4 $ & $110 \pm 3$ \\
  Si K$\beta$ & $11.5 \pm 1.8$ & $7.64 \pm 0.89$ \\
  S K$\alpha$ & $33.3 \pm 2.8$ & $25.4 ^{+1.9}_{-2.0}$ \\
  Ar K$\alpha$ & $2.94 \pm 1.02$ & $3.56 \pm 0.88$ \\
  Ca K$\alpha$ & $0.992^{+0.71}$ & $1.08 \pm 0.46$ \\
  Fe K$\alpha$ & $4.60^{+2.00}_{-1.94}$ & $3.32^{+0.66}_{-0.65} $ \\

  \cutinhead{Line Centroids (keV)}
  Si K$\alpha$ & $1.833^{+0.003}_{-0.002}$ & $1.834 \pm 0.002$ \\
  Si K$\beta$ & $2.146^{+0.019}_{-0.014}$ & $2.161 \pm 0.014$ \\
  S K$\alpha$ & $2.415^{+0.006}_{-0.007}$ & $2.422^{+0.006}_{-0.005}$ \\
  Ar K$\alpha$ & $3.064^{+0.027}_{-0.024}$ & $3.036 \pm 0.042$ \\
  Ca K$\alpha$ & $3.737^{+0.072}_{-0.060}$ & $3.811^{+0.032}_{-0.033}$ \\
  Fe K$\alpha$ & $6.572_{-0.208}$ & $6.440^{+0.030}_{-0.029}$ \\

  \cutinhead{Goodness-of-Fit}
  $\chi^{2}$/dof & $45.1/64$ & $135.0/130$ \\
  \enddata

 \tablecomments{The limits given are the formal $90\%$ confidence ranges $(\Delta \chi^{2}=2.706)$.}

 \tablenotetext{a}{Upper limit to the flux in the O K$\alpha$ blend (0.55-0.58 keV) assuming $N_{H}=0.076 \times
   10^{22}\,\mathrm{cm^{-2}}$} 

 \tablenotetext{b}{Flux in the 0.8-1.2 keV band assuming $N_{H}=0.07 \times 10^{22}\,\mathrm{cm^{-2}}$, including
   unresolved contributions from Ne K$\alpha$ and other lines.}

\end{deluxetable}

\begin{deluxetable}{lcc}
  \tablewidth{0pt}
  \tabletypesize{\scriptsize}
  \tablecaption{Line flux ratios in the SNR \of \label{tab-2}}
  \tablecolumns{3}
  \tablehead{
    \colhead{} &
    \colhead{\ch} &
    \colhead{\xmm}\\
    \colhead{Line Ratio} &
    \colhead{ACIS-S} &
    \colhead{EPIC-pn}
  }
    
  \startdata
  O K$\alpha$/Si K$\alpha$ & $< 3.2$ & $< 3.3$ \\
  Fe L/Si K$\alpha$ & $9.6$ & $8.7$ \\
  Si K$\beta$/Si K$\alpha$ & $0.093 \pm 0.018$ & $0.069 \pm 0.010$ \\
  S K$\alpha$/Si K$\alpha$ & $0.27 \pm 0.03$ & $0.23 \pm 0.02$ \\
  Ar K$\alpha$/Si K$\alpha$ & $0.024 \pm 0.009$ & $0.032 \pm 0.009$ \\
  Ca K$\alpha$/Si K$\alpha$ & $0.0081 \pm 0.0060$ & $0.0098 \pm 0.0044$\\
  Fe K$\alpha$/Si K$\alpha$ & $0.037 \pm 0.017$ & $0.030 \pm 0.007$
  \enddata

  \tablecomments{For simplicity, symmetric confidence ranges have been calculated taking the largest deviations from
    the fitted values in each case.}

\end{deluxetable}

The values listed in Tables \ref{tab-1} and \ref{tab-2} merit a few comments. All the line centroids and line flux
ratios, as well as the power law indices and normalizations, are consistent in both data sets within the statistical
uncertainties. However, some puzzling differences between the \ch\ ACIS-S and the \xmm\ EPIC-pn spectra can be
appreciated both in the Tables and in Figures \ref{fig-3} and \ref{fig-4}. The fluxes of the three brightest lines are
statistically inconsistent, with deviations of $11\%$ for Si K$\alpha$, $24\%$ for S K$\alpha$, and a very significant
$34\%$ for Si K$\beta$. The upper limits for the O K$\alpha$ emission and the fluxes in the broad Fe L band also show
disagreements of $9\%$ and $18\%$. Some of these differences are larger than the cross-calibration uncertainty between
\ch\ and \xmm, which is $15\%$ in normalisation between 0.8 and 2 keV (M.Stuhlinger et al., document
XMM-SOC-CAL-TN-0052, Issue 3.0, January 2006). The discrepancies certainly deserve some further investigation, but they
are of no consequence for the scientific objectives of this paper, for two reasons that will become clear along
\S~\ref{sec:Modeling}. First, we only model parameters that are consistent in both data sets. And second, the tolerance
ranges that we use to compare models and data are much larger than the differences between ACIS-S and EPIC-pn that we
have discussed here, and could easily accomodate the EPIC-MOS data sets as well.

\section{SPECTRAL MODELING}
\label{sec:Modeling}

\subsection{Method, Parameters, and Strategy}
\label{sub:Methods}

To compare our grid of DDT models to the observations of SNR \of, we have generated synthetic X-ray spectra for the
shocked ejecta emission in our hydrodynamic simulations following the methods presented in
\citet{badenes03:xray,badenes05:xray}, with updated atomic data \citep{badenes06:tycho}, and including radiative and
ionization losses with an isobaric approximation as described in \citet{badenes07:outflows}. For a given Type Ia SN
explosion model, the synthetic spectra are controlled by three variables only: AM density $\rho_{AM}$, SNR age $t$, and
$\beta$. The parameter $\beta$ represents the amount of collisionless electron heating at the reverse shock, and is
defined as the ratio of specific internal energies in electrons and ions at the shock transition. It can take values
between $\beta_{min}$, which represents mass proportional heating, and 1, which represents full equilibration \citep[for
more details, see \S~2.2 in][]{badenes05:xray}. We have performed simulations in a grid of seven $\rho_{AM}$ values ($5
\times 10^{-26}$, $10^{-25}$, $2 \times 10^{-25}$, $5 \times 10^{-25}$, $10^{-24}$, $2 \times 10^{-24}$, and $5 \times
10^{-24}\,\mathrm{g \, cm^{-3}}$) and three $\beta$ values ($\beta_{min}$, 0.01, and 0.1). A similar, less extended grid
of synthetic X-ray spectra based on DDT models was presented and discussed in \citet{badenes05:model_grid}.

The ability of our synthetic spectra to reproduce SNR observations is limited by the quality of the atomic data in the
spectral code. In the particular case of \of, several important issues arise due to the low ionization state of the
plasma. At ionization states below He-like, the atomic data for the K$\alpha$ blends of the major elements are
reasonably complete, but deficiencies in K-shell transitions from higher levels should be expected in all
elements. Atomic data in the Fe L complex are notably deficient, and altogether absent for ionization stages below
Fe$^{+16}$ (Ne-like Fe). The strongest Fe L line in the \ch\ HETG spectrum (at $\sim 0.73$ keV) can be associated with
Ne-like Fe, and there are no signs of a significant contribution from lower ionization stages (Hughes et al., in
preparation). Nevertheless, Fe L emission in the synthetic spectra should always be considered with caution as a matter
of principle.

Given the limitations of the synthetic spectra listed above and the issues with the X-ray data described in
\S~\ref{sub:Spectral-Fits}, we have chosen to focus our efforts on modeling only fundamental quantities for which we can
trust both models and observations. These include the centroids of the three brightest line blends (Si K$\alpha$, S
K$\alpha$, and Fe K$\alpha$), and the O K$\alpha$/Si K$\alpha$, Fe L/Si K$\alpha$ and Fe K$\alpha$/Si K$\alpha$ line
flux ratios, noting again that special care must be taken in the case of the Fe L/Si K$\alpha$ ratio. As we shall see,
the combination of these parameters can constrain the kinetic energy of the SN explosion in the framework of the DDT
models quite well. We will begin by using the DDT models at the nominal age of 400 yr in \S~\ref{sub:DDT-400}, and then
discuss variations of age in \S~\ref{sub:DDT_Otherages} and alternative models in \S~\ref{sub:Other_Models}. In
\S~\ref{sub:Sp_Int} we will go beyond our selection of diagnostic parameters to evaluate the performance of the models
across the entire spectral range of \xmm.

\subsection{DDT Models at Nominal Age (t=400 yr)}
\label{sub:DDT-400}

A comparison between the observed line flux ratios and centroids and the predictions of the synthetic spectra generated
by the DDT models in our grid at $t=400$ yr is presented in Figure \ref{fig-5}. In these plots, the observed values
(vertical solid lines) are the averages of the \ch\ ACIS-S and \xmm\ EPIC-pn values, except in the case of the Fe
K$\alpha$ line, where only the \xmm\ EPIC-pn data are taken into account. We note that the largest difference between
ACIS-S and EPIC-pn in the relevant parameters corresponds to the the Fe L/Si K$\alpha$ ratio, and is only
$9\%$. Tolerance ranges around the observed values are highlighted with striped regions: a factor 2 above and below for
line flux ratios (except for the O K$\alpha$/Si K$\alpha$ ratio, where only the upper limit is constrained), $1 \%$ for
the Si K$\alpha$ and S K$\alpha$ centroids, and $0.5 \%$ for the Fe K$\alpha$ centroid. These tolerance ranges are
rather large, mostly because of systematic uncertainties that are very hard to quantify \citep[see \S~5.2 in][for a
discussion]{badenes06:tycho}, and will only be used to provide some guidance in the comparisons between models and
data. For each DDT model, the values of $\rho_{AM}$ that satisfy the dynamical constraints from \S~\ref{sec:Dynamics} at
$t=400$ yr are indicated by horizontal dotted lines.

The combination of the O K$\alpha$/Si K$\alpha$ and Fe K$\alpha$/Si K$\alpha$ flux ratios clearly favors models with low
O and high Fe emission. Only the DDT models with higher kinetic energies (and hence high Fe and low O yields), DDTa and
DDTc, can reproduce both observables at the same time. When the dynamical constraints are brought into play, model DDTa
(for $\rho_{AM}=10^{-24}\,\mathrm{g\,cm^{-3}}$ and $\beta$ between 0.01 and 0.1) performs better than model DDTc,
because of its lower O K$\alpha$ flux. The Fe L/Si K$\alpha$ ratio also favors model DDTa over DDTc, although we stress
that comparisons based on this parameter must be of a qualitative sort. The high Fe L/Si K$\alpha$ values found at low
$\rho_{AM}$ in models DDTc, DDTe and DDTg, for instance, are due mostly to Ne K emission, not to L-shell lines from
shocked Fe. The line centroids plotted in the right hand panels of Figure \ref{fig-5} support the choice of model DDTa,
but we note that the Fe K$\alpha$ centroid is slightly underpredicted at the $\rho_{AM}$ imposed by the dynamical
constraints.

\begin{figure}

  \centering
 
  \includegraphics[angle=90,scale=0.9]{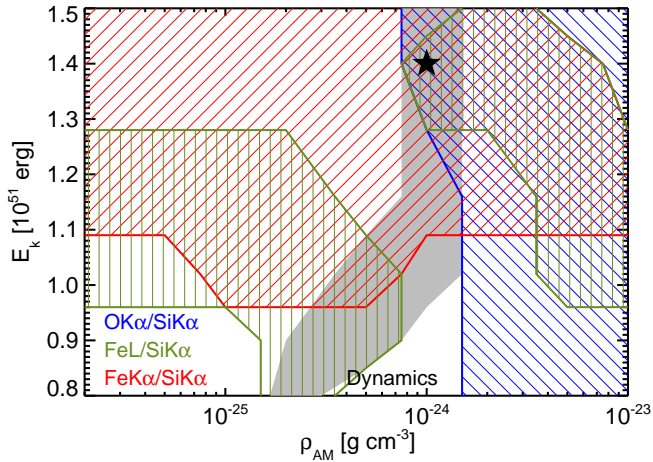}

  \caption{Global view of the performance of the DDT models (mapped onto an $E_{k}$ axis) with varying $\rho_{AM}$ at
    $t=400$ yr. The striped regions represent the areas of the parameter space where the models can reproduce the
    observed O K$\alpha$/Si K$\alpha$ (blue), Fe L/Si K$\alpha$ (green), and Fe K$\alpha$/Si K$\alpha$ (red) line flux
    ratios for an optimum value of $\beta$. The dynamical constraints on $\rho_{AM}$ are overlaid in gray. The star
    indicates the position of model DDTa at $\rho_{AM}=10^{-24}\,\mathrm{g\,cm^{-3}}$, where all the constraints
    overlap. \label{fig-6}}

\end{figure}

The interplay between the constraints from the different line flux ratios with varying $\rho_{AM}$ can be seen more
clearly in Figure \ref{fig-6}, where the sequence of DDT models has been mapped onto an $E_{k}$ axis. The O K$\alpha$/Si
K$\alpha$ ratio can be reproduced by any model, provided that $\rho_{AM}$ is high enough to ionize most O beyond the
He-like stage at $t=400$ yr. On the other hand, the Fe K$\alpha$/Si K$\alpha$ ratio can only be reproduced by models
with high $E_{k}$, where enough Fe has been shocked at $t=400$ yr. These two flux ratios, together with the dynamical
constraints clearly single out model DDTa with $\rho_{AM}=10^{-24}\,\mathrm{g\,cm^{-3}}$. The Fe L/Si K$\alpha$ ratio
confirms the choice, but we note again that the allowed region at low $\rho_{AM}$ for this parameter is spurious. This
plot highlights the remarkable agreement between the dynamical and X-ray spectral constraints on $\rho_{AM}$ for the
energetic DDT models, specially considering the large dynamic range of the diagnostic quantities shown in Figure
\ref{fig-5}. This is not a trivial coincidence, and it is not observed in other explosion models, as we shall see in
\ref{sub:Other_Models}. It indicates that the fundamental properties of the DDT models (exponential density profile,
stratified ejecta) are well suited for SNR \of.

\subsection{DDT Models at Different Ages}
\label{sub:DDT_Otherages}

\begin{figure*}

  \centering
  
  \includegraphics[angle=90,scale=0.8]{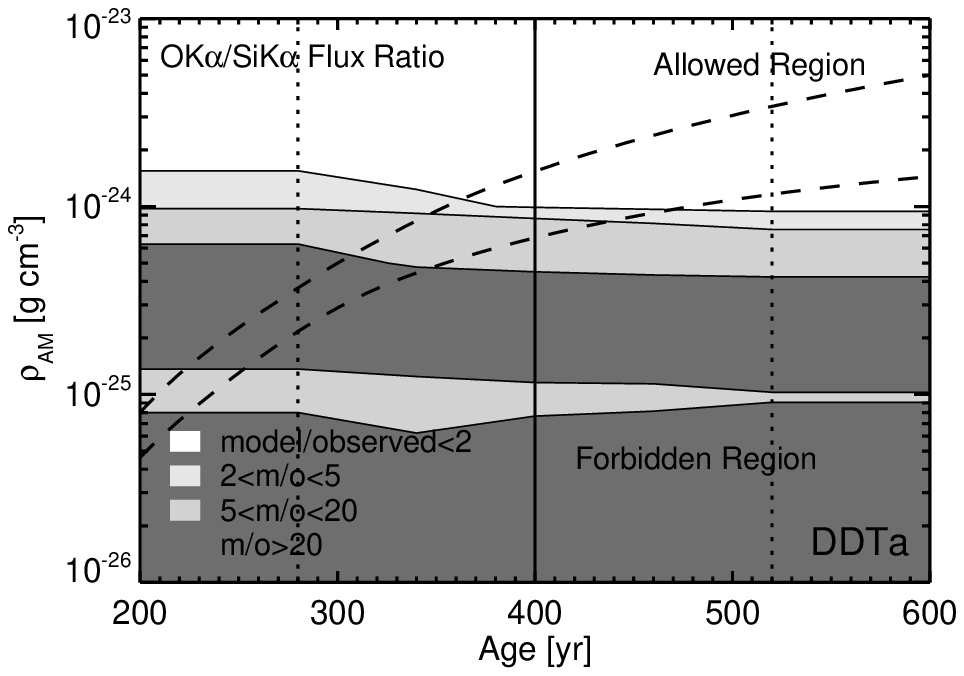}
  \includegraphics[angle=90,scale=0.8]{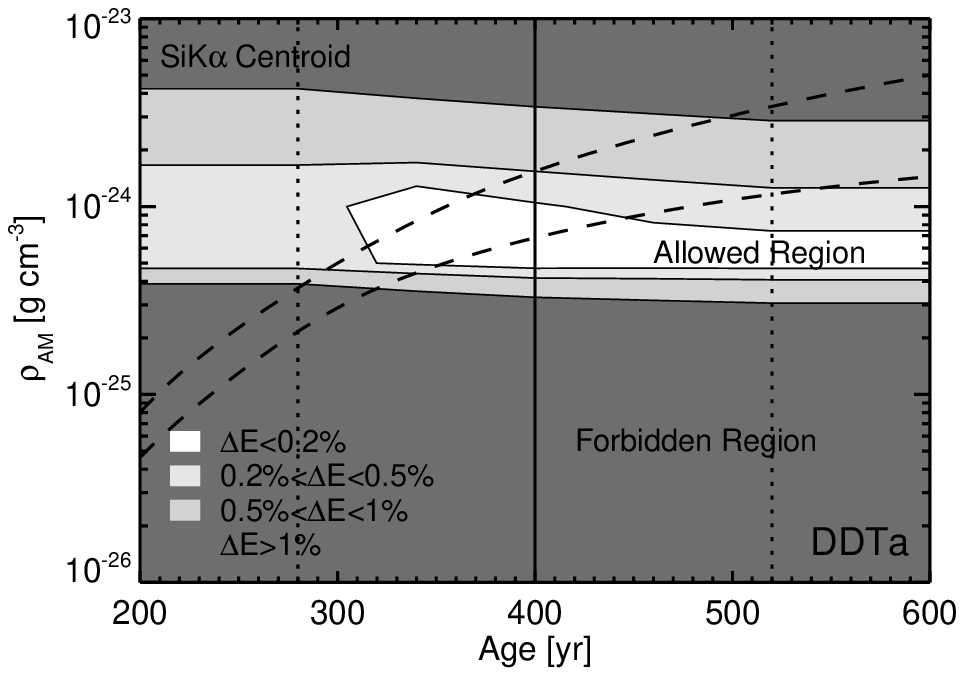}

  \caption{Comparison between the observations and the values predicted by the synthetic spectra produced from model
    DDTa for the O K$\alpha$/Si K$\alpha$ flux ratio (left panel) and the centroid of the Si K$\alpha$ blend (right
    panel) as a function of $\rho_{AM}$ and $t$. The black dashed plots are the dynamical constraints from the SNR
    radius from Figure \ref{fig-2}. The age estimate from the light echoes is represented with the vertical solid and
    dotted lines. \label{fig-7}}

\end{figure*}

The uncertainties in the age estimate from R05 ($\pm120$ yr around the nominal age of 400 yr) stem from the unknown
three-dimensional structure of the dust that is reflecting the light echo. The SNR dynamics are consistent with this
range, but they impose a tight correlation between $t$ and $\rho_{AM}$, as shown in \S~\ref{sec:Dynamics}. Now we shall
see that the properties of the X-ray emission from \of\ can substantially narrow down the age estimate, making large
deviations from $t=400$ yr seem unlikely. Higher ages require higher values of $\rho_{AM}$ in order to reproduce the FS
radius, and \textit{vice versa} (see Figure \ref{fig-2}). These changes in $\rho_{AM}$ lead to major differences in the
ionization timescale $\tau = \intop n_{e}dt$ of the shocked ejecta, which rapidly drive several spectral parameters out
of agreement with the observations. Two examples are presented in Figure \ref{fig-7}, where the O K$\alpha$/Si K$\alpha$
flux ratio and the centroid of the Si K$\alpha$ blend are compared with the observed values across the ($\rho_{AM}$,
$t$) parameter space for model DDTa. At ages below 400 yr, most of the O is in the He-like ionization stage for the
$\rho_{AM}$ values imposed by the FS radius. This makes the O K$\alpha$ emission too strong, with O K$\alpha$/Si
K$\alpha$ ratios in excess of 5 times the observed value. Above 400 yr, the O K$\alpha$/Si K$\alpha$ ratio is in the
allowed region, but then the He-like ion of Si becomes dominant in the plasma, pushing the centroid toward higher
energies that are incompatible with the value measured by \ch\ and \xmm. These comparisons are less favorable for the
other DDT models, because their higher O content and lower $E_{k}$ tend to increase the disagreement between the
dynamical and spectral constraints.

One last constraint on the age of \of\ comes from the lack of recorded historical evidence for a recent SN in the LMC. A
normal to bright Type Ia SN at a distance of 50 kpc would remain visible for several months, with an apparent visual
magnitude around $-$1.0 at maximum light \citep[the average extinction towards the LMC is 0.3
mag,][]{imara07:LMC_extinction}. Such an event would be hard to miss by even the most inattentive observers. European
exploration of the southern hemisphere was well under way 400 years ago, led by skilled navigators who relied heavily on
astronomy and monitored the night sky whenever possible. Pieter Dirkszoon Keyser and Frederik de Houtman mapped the
southern sky (including Mensa and Doradus, which encompass the LMC) from Java between 1595 and 1597, noting the
positions of all the stars brighter than 5$\mathrm{^{th}}$ magnitude. In this island, the Dutch colony of Batavia
(present day Jakarta, at $6 ^\circ 16'$ S latitude) and the English colony of Bantam became permanent around 1600. In
South America, several Spanish settlements were thriving decades before that, including Lima (founded 1535, $12 ^\circ
12.6'$ S) and Buenos Aires (permanent since 1580, $34 ^\circ 36'$ S), from which the LMC is circumpolar. Even allowing
for some incompleteness in the historical records, a spectacular astronomical event like a bright LMC supernova should
have left some kind of trace if it happened at any time later than the beginning of the 17$\mathrm{^{th}}$ century.

Taken together, the light echoes from the SN, the dynamic and spectral properties of the SNR, and the historical
considerations, suggest an age very close to $400$ yr for the birth event of \of. The uncertainty around this value is
hard to quantify, but deviations larger than a few decades seem unlikely, specially towards younger ages.

\subsection{Other Models}
\label{sub:Other_Models}

\begin{figure}

  \centering
 
  \includegraphics[scale=0.65]{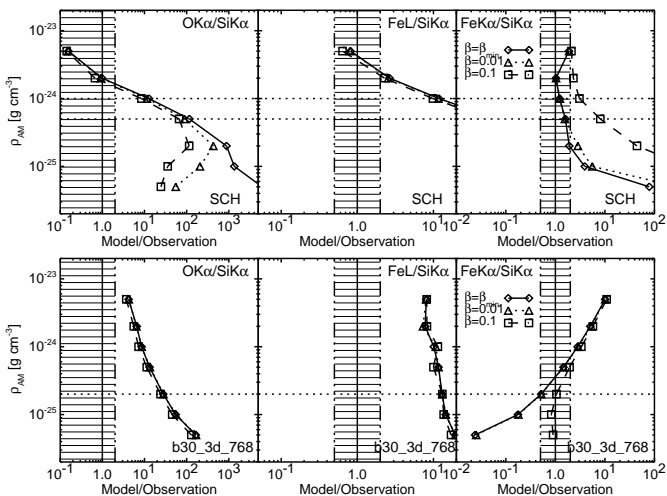}

  \caption{Fundamental line flux ratios as a function of $\rho_{AM}$ form models SCH (top panels) and b30\_3d\_768 (bottom
    panels) at $t=400$ yr. All labels and plots are as in Figure \ref{fig-5}. \label{fig-8}}

\end{figure}

We have seen that one-dimensional DDT models can reproduce the fundamental properties of the ejecta emission in \of\
with remarkable accuracy. It is outside the scope of this paper to perform an exhaustive exploration of other Type Ia
explosion paradigms like \citet{badenes06:tycho} did for the Tycho SNR, but for the sake of completeness we discuss here
some results obtained using other models. In Figure \ref{fig-8} we present the line flux ratios at $t=400$ yr predicted
by a sub-Chandrasekhar explosion \citep[model SCH from][]{badenes03:xray} and a well-mixed 3D deflagration \citep[model
b30\_3d\_768 from][]{travaglio04:3D}. In each case, we have overlaid the dynamical constraints on $\rho_{AM}$ obtained
with the procedure explained in \S~\ref{sec:Dynamics}. The sub-Chandrasekhar model clearly shows that the overlap
between spectral and dynamical constraints found in the energetic DDT models is not trivial, and indeed does not happen
for the ejecta structure of this edge-lit WD detonation. The well-mixed 3D deflagration model has more severe problems:
the presence of Fe and O everywhere in the ejecta leads to a systematic overprediction of the O K$\alpha$/Si K$\alpha$
and Fe L/Si K$\alpha$ flux ratios at all values of $\rho_{AM}$.

\subsection{Spatially Integrated Spectra}
\label{sub:Sp_Int}

\begin{figure*}

  \centering
 
  \includegraphics[angle=90,scale=0.8]{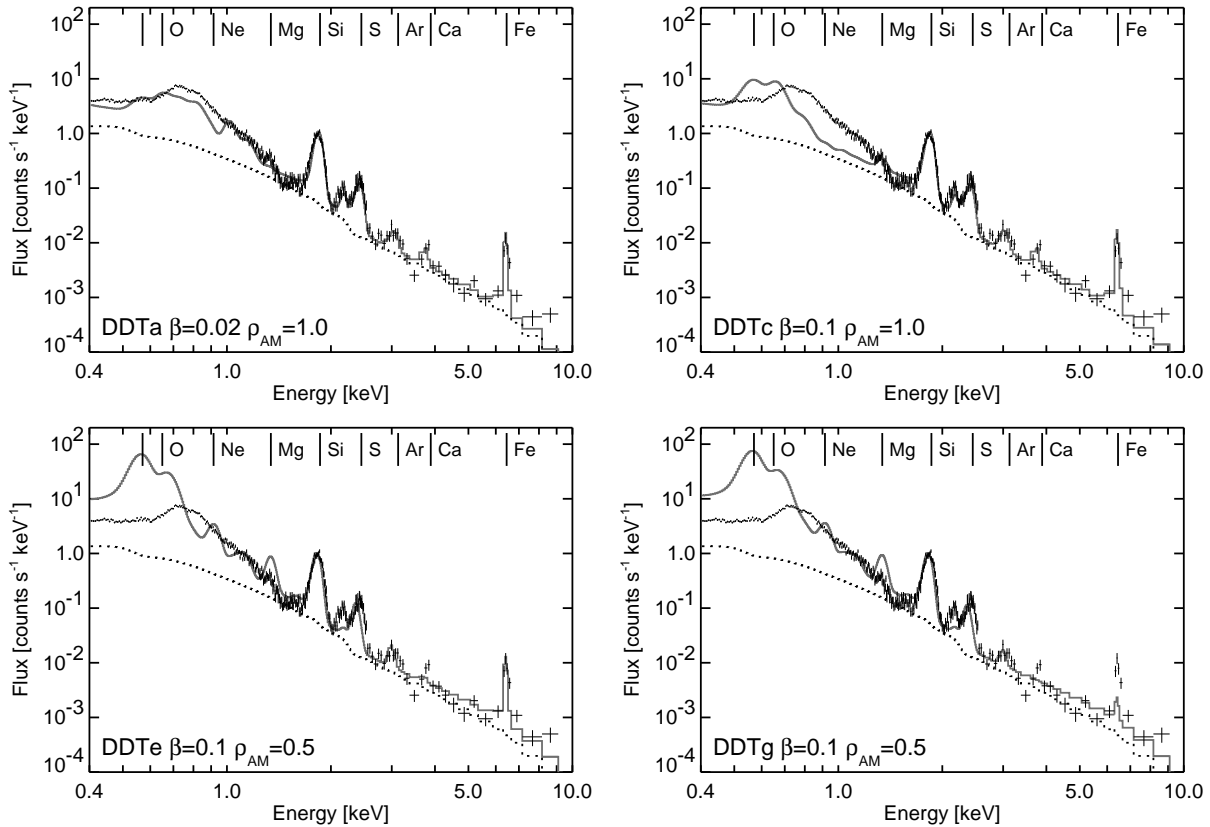}

  \caption{Comparison between the \xmm\ EPIC-pn data set and the DDT model spectra at $t=400$ yr for the most favorable
    values of $\beta$ and $\rho_{AM}$ (in units of $10^{-24}\,\mathrm{g~cm^{-3}}$) within the dynamical
    constraints. Each ejecta model has been normalized to match the flux in the Si K$\alpha$ blend, and then a power law
    model with the parameters listed in Table \ref{tab-1} has been added. Absorption has been fixed at $N_{H}=0.07
    \times 10^{22}\,\mathrm{cm^{2}}$. The position of the principal emission lines (K$\alpha$ and Ly$\alpha$ for O,
    K$\alpha$ for all other elements) has been indicated for clarity. \label{fig-9}}

\end{figure*}

In the previous Sections we have taken the approach of focusing on a few parameters that can be both reliably extracted
from the observations \textit{and} confidently modeled with the existing spectral codes. This is a more meaningful and
robust way of comparing synthetic spectra generated with HD+NEI models to data than the conventional method of
$\chi^{2}$ fitting across the entire X-ray band \citep[for discussions,
see][]{badenes05:xray,badenes06:tycho}. Nevertheless, it is always instructive to study the performance of the spectral
models from a more global point of view. In Figure \ref{fig-9}, we plot the synthetic spectra of the DDT models at
$t=400$ yr for the best value of $\rho_{AM}$ within the dynamical constraints, together with the \xmm\ EPIC-pn data
set. In each model, the value of $\beta$ has been chosen to give the best approximation to the Fe K$\alpha$/Si K$\alpha$
flux ratio: $\beta=0.02$ for model DDTa and $\beta=0.1$ for models DDTc, DDTe, and DDTg. No spectral fitting of any kind
has been done. Instead, the ejecta models have been normalized to match the Si K$\alpha$ line and then a power law with
the parameters determined in \S~\ref{sub:Spectral-Fits} ($\alpha=3.42$, $Norm=3.5 \times 10^{-4} \, \mathrm{photons \,
  cm^{-2}} \, s^{-1}$) has been added for the continuum. Absorption has been set to the fiducial $N_{H}=0.07 \times
10^{22}\,\mathrm{cm^{2}}$ in all cases. By foregoing any spectral fitting, we can more easily compare the properties of
the ejecta models to the data and to each other.

The performance of our model of choice, DDTa ($\rho_{AM}=10^{-24}\,\mathrm{g\,cm^{-3}}$, $\beta=0.02$) is very good at
all energies, albeit with some problems in localized areas. The flux in the Fe L complex is somewhat low, and the shape
shows deviations from the observed spectrum. These discrepancies might be due mostly to the limitations in the atomic
data discussed in \S~\ref{sub:Methods}. Other problems are probably related to the explosion model, such as the deficit
in the Mg K$\alpha$ flux and the low centroid of the Ca K$\alpha$ blend\footnote{The same two issues were noted by
  \citet{badenes06:tycho} in comparisons between the synthetic spectrum from model DDTc ($t=430$ yr,
  $\rho_{AM}=2\times10^{-24}\,\mathrm{g\,cm^{-3}}$, $\beta=0.03$) and the X-ray emission from the Tycho SNR. This
  suggests that the Mg content in our DDT models might be underestimated, and the distribution of Ca in the ejecta might
  be biased towards low densities.}. The degradation of the synthetic spectra along the sequence of DDT models with
decreasing $E_{k}$ (or $\rho_{tr}$) is plain to see in Figure \ref{fig-9}, specially in the behavior of the Fe and O
emission. Normalization distances can be calculated for each model, yielding 36 kpc (DDTa), 51 kpc (DDTc), 23 kpc
(DDTe), and 29 kpc (DDTg). Given the coarseness of our $\rho_{AM}$ grid, values within $\sim50 \%$ of the distance to
the LMC can be considered satisfactory. A more accurate study of the normalization distances cannot be justified without
the inclusion of multidimensional effects in the HD+NEI simulations \citep[see \S~8.1 in][]{badenes06:tycho}.

We conclude this Section with a brief mention of the continuum emission. Detailed modeling of this component of the
X-ray spectrum is outside the scope of the present work, but our HD+NEI calculations can help clarify the issue of its
origin raised by WH04. At high energies ($>2$ keV), the flux of the thermal AM emission in our HD+NEI models (for
$\rho_{AM}=10^{-24}\,\mathrm{g\,cm^{-3}}$ and $t=400$ yr at a distance of 50 kpc) is a factor 6 below the power law
models listed in Table \ref{tab-1}. This is a strong indication that the origin of the continuum emission in \of\ is
indeed nonthermal.


\section{DISCUSSION AND CONCLUSIONS}
\label{sec:Disc-Concl}

The light echo observations of R05 and R07 have turned \of\ into a unique object: the only X-ray bright SNR whose Type
Ia origin is confirmed and whose explosion energy can be estimated based on spectroscopic information from its parent
supernova. Motivated by these groundbreaking observations, we have re-examined the dynamics and X-ray spectrum of \of,
applying the HD+NEI modeling techniques introduced in \citet{badenes03:xray} and \citet{badenes05:xray} to \ch\ and
\xmm\ observations. We have based our analysis on a grid of one-dimensional DDT explosions, taking advantage of the
relationship between ejecta structure and $E_{k}$ in these models to place the birth event of \of\ in the sequence of
dim to bright Type Ia SNe. Our conclusions are in excellent agreement with the light echo results of R07: SNR \of\ was
originated by a bright, highly energetic Type Ia event of the subtype often referred to as SN 1991T-like. We thus
present, for the first time, a scenario where the dynamics and X-ray spectrum of a Type Ia SNR form a consistent picture
with the spectroscopy of its parent supernova.

Our preferred Type Ia explosion model, DDTa, has a kinetic energy of $1.40 \times 10^{51}$ erg, and synthesizes $1.03 \,
\mathrm{M_{\odot}}$ of Fe, $0.09 \, \mathrm{M_{\odot}}$ of Si, $0.07 \, \mathrm{M_{\odot}}$ of S, $0.04 \,
\mathrm{M_{\odot}}$ of O, and $0.14 \, \mathrm{M_{\odot}}$ of other elements (mostly Ar and Ca). The peak bolometric
luminosity of this model is $\log L_{\mathrm{bol}} (\mathrm{ergs~s^{-1}})=43.39$, and the amount of $^{56}$Ni in the
ejecta before nuclear decays is $0.97\,\mathrm{M_{\odot}}$. These values compare very well with the estimates for SN
1991T: $\log L_{\mathrm{bol}}=43.36$ \citep{contardo00:bolometric_light_curves_Ia};
$\mathrm{M_{^{56}Ni}}=0.87\,\mathrm{M_{\odot}}$ \citep[][UVOIR method]{stritzinger06:consistent_estimates_56Ni_Ias},
$0.94\,\mathrm{M_{\odot}}$ \citep[][B-band lightcurve
method]{stritzinger06:consistent_estimates_56Ni_Ias,mazzali07:zorro}.  The ejecta density and chemical composition
profiles in model DDTa ($\rho_{AM}=10^{-24}\,\mathrm{g \, cm^{-3}}$, $\beta=0.02$) can reproduce very well both the
dynamics (FS radius and velocity) and the fundamental properties of the X-ray spectrum (O, Si, S, and Fe emission) of
\of\ at an age of 400 yr. Several dynamical, spectral, and historical arguments indicate that the SNR age cannot be very
different from this value, which also coincides with the completely independent estimate derived by R05 from the
geometry of the light echo.

These results, together with our previous work on the Tycho SNR \citep{badenes06:tycho}, constitute firm evidence that
the phenomenological one-dimensional DDT models, which have proved so successful in explaining the light curves and
spectra of Type Ia SNe, are equally capable of reproducing the fundamental properties of the X-ray emission from young
Type Ia SNRs. In particular, it is possible to use HD+NEI simulations based on these DDT models to estimate the kinetic
energy (and ultimately the brightness) of a Type Ia SN explosion from the X-ray spectrum of its SNR. Hundreds of years
after the SN explosion, the memory of the cataclysmic event persists in the X-rays from its SNR, opening a window onto
ages past. This enables us to explore the relationship between individual dim/bright Type Ia SNe and their immediate
surroundings (presence or absence of stellar formation, local metallicities, etc.) with much greater detail than is
available to extragalactic studies \citep[see][]{prieto07:SN_Progenitors_Metallicities}. In the context of SNR research,
having a good estimate for $E_{k}$ can help to build better models for the impact of CR acceleration on the SNR
dynamics. HD+NEI simulations have clearly become a powerful and flexible tool to study the relationship between SNRs and
their parent SNe. For interested readers, the synthetic X-ray spectra from our models are available from the authors
upon request.

We conclude with a reminder that much is left to do in the study of Type Ia SNe and their SNRs in general, and of \of\
in particular. Ongoing direct measurements of the ejecta and FS expansion using grating observations and proper motion
studies should improve our knowledge of the dynamics of this SNR in the near future. Further work on the X-ray emission
should take the asymmetries of the object into account, in particular the enhanced Fe knots found by WH04. The
one-dimensional models presented here can only sample the average or bulk properties of the shocked
ejecta. Observational evidence for moderate asymmetries \citep{fesen07:SN1885,gerardy07:SNIa_midIR} and clumping
\citep{leonard05:spectropolarimetric_diversity_SNIa,wang07:SNIa_spectropolarimetry} in the ejecta of Type Ia SNe is
widespread, and SNR studies offer a unique opportunity to study these effects. This might provide crucial constraints
for the multi-dimensional simulations of physically-motivated DDT explosions currently under way
\citep{bravo06:PRD,plewa07:DFD,jordan07:GCD_3D}

\acknowledgements The authors would like to thank Armin Rest, Tom Matheson, Jos\'{e} L. Prieto, and the rest of the
SuperMACHO team for their stimulating work on the light echoes from \of\ and for discussing several details of their
results in advance of publication. CB also wishes to thank Edward van den Heuvel for his engaging presentation on the
astronomical observations performed by Dutch explorers in the 17$\mathrm{^{th}}$ century. Support for this work was
provided by the National Aeronautics and Space Administration through Chandra Postdoctoral Fellowship Award Number
PF6-70046 issued by the \ch\ X-ray Observatory Center, which is operated by the Smithsonian Astrophysical Observatory
for and on behalf of the National Aeronautics and Space Administration under contract NAS8-03060. Additional support was
provided by the National Science Foundation under Grant No. PHY05-51164 to KITP-UCSB and by \ch\ grant GO7-8068X to
Rutgers University. E. B. has received support from the DURSI of the Generalitat de Catalunya and the Spanish DGICYT
grants AYA 2004-06290-C02-02 and AYA 2005-08013-C03-01.



\clearpage

\appendix
\section{SOME REFLECTIONS ON THE RELEVANCE OF SURREALISM TO ASTROPHYSICS}

\begin{figure*}
  \centering
 
  \includegraphics[scale=0.7]{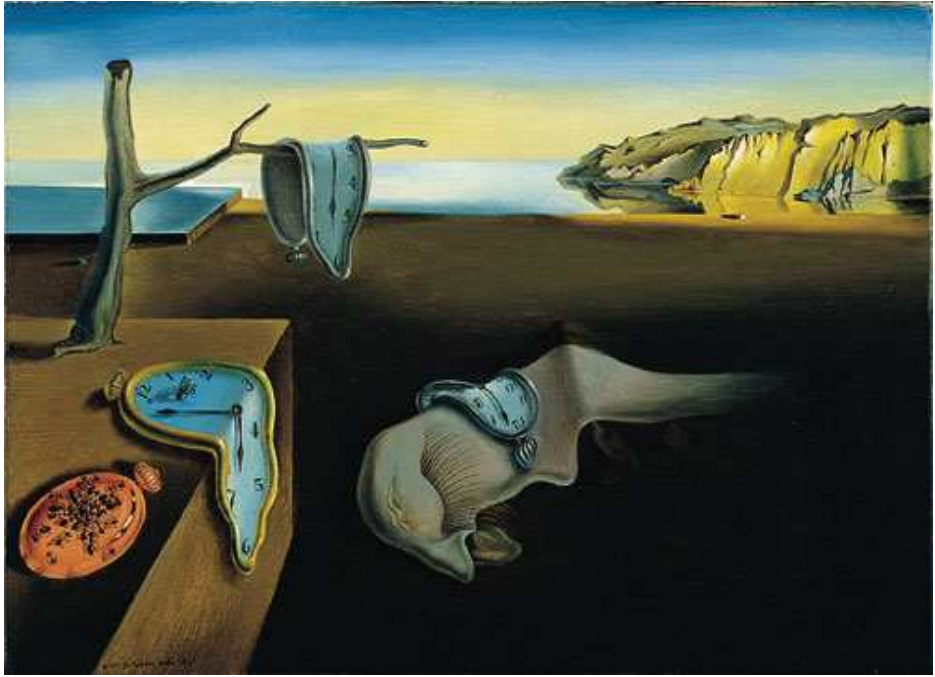}

  \vspace{1cm}

  \caption{`The Persistence of Memory', by Salvador Dal\'{i}, 1931. Oil on canvas, Museum of Modern Art, New York
    City. This painting by the genial Catalan artist represents the passage of time, and is very relevant to our work
    on SNR \of\ in a convoluted, surreal way. Dal\'{i} referred to the vaguely human figure in the center as the
    `paranoiac-critical Camembert', for whatever that is worth.  \label{fig-10}}

\end{figure*}

The meaning of Salvador Dal\'{i}'s painting \textit{The Persistence of Memory} (Figure \ref{fig-10}) has been discussed
at length elsewhere (see f.i. \textit{MoMA Highlights}, New York: The Museum of Modern Art). The most common
interpretation is that the famous `soft clocks' are a metaphor for the passage of time, and the ants represent the
inevitable consequence thereof: corruption of the flesh. Other interpretations are definitely possible, and equally
valid - this being, after all, surrealism. Dal\'{i} once commented that the idea of the soft clocks came to him from the
observation of a piece of Camembert cheese melting in the summer heat. Be that as it may, if the soft clocks \textit{do}
represent the passage of time, it is easy to draw an analogy between \textit{The Persistence of Memory} and young SNRs
like \of. The X-ray observations of these objects open a window onto ages past, telling us what processes went on during
the supernova explosions that originated them, and further back in time, how the supernova progenitor modified its
surroundings. Just like a piece of melting Camembert cheese.

\end{document}